\documentclass{ws-mpla}
\usepackage{color}
\usepackage{graphicx}% Include figure files
\usepackage{dcolumn}% Align table columns on decimal point
\usepackage{bm}% bold math
\usepackage{epsfig}
\begin{document}

\markboth{Xiao-Gang He} {CKM Phase and Spontaneous $CP$ Violation}

%%%%%%%%%%%%%%%%%%%%% Publisher's Area please ignore %%%%%%%%%%%%%%
%\catchline{}{}{}{}{}
%%%%%%%%%%%%%%%%%%%%%%%%%%%%%%%%%%%%%%%%%%%%%%%%%%%%%%%%%%%%%%%%%%%

\begin{center}
{\Large {\bf CKM Phase and Spontaneous $CP$ Violation}\footnote{hexg@phys.ntu.edu.tw}}
\end{center}

\begin{center}
{Xiao-Gang He}\\
\vspace*{0.3cm}

Department of Physics and Center for Theoretical Sciences\\
National Taiwan University, Taipei, Taiwan
\end{center}
%\maketitle

%\pub{Received (Day Month Year)}{}%Revised (Day Month Year)}

%\begin{abstract}
\begin{center}
\begin{minipage}{12cm}

\noindent Abstract:\\
The Standard Model for CP violation,
the CKM model, works very well in explaining all laboratory experimental data.
However, this model does not address the question that where it comes from.
The origin of CP violation is still a mystery.
In this talk I discuss a model\cite{model} addressing this problem
in which the CP violating phase in the CKM matrix is identical to
the phase in the Higgs potential resulting from spontaneous CP
violation.

%\end{abstract}
\end{minipage}

\end{center}
%\ccode{PACS Nos.: include PACS Nos.}

Since the discovery of parity (P) violation in weak
interactions\cite{parity} by T.D. Lee and C.-N. Yang in 1956,
great progresses have been made in many ways. Parity violation is
now understood to be due to $V-A$ current interaction in weak
interactions. We have a successful theory for electroweak
interactions - the Standard Model (SM)\cite{sm}. Tremendous progresses in
understanding discrete space-time symmetries, the Parity $P$,
the time reversal $T$ and charge conjugation $C$, have also been
made over the last 50 years or so. In 1964, CP was found to be
violated in neutral kaon decays into two and three
pions\cite{cpv}. Since then CP violation have been observed in
kaon decay amplitude, and in B meson decays in recent
years\cite{pdg}. Now we have a very successful model for CP
violation, the CKM model\cite{km}. $T$ violation has also been
established in kaon decays\cite{pdg}. There is no evidence of
violation for the combined symmetry CPT.

It was realized in 1973 by Kobayashi and Maskawa\cite{km} that in
the minimal SM if there is a miss-match between weak and mass
eigenstates of quarks in the interaction with the weak gauge boson
$W$ and Higgs boson, it is possible to have CP violation.  In the mass
eigenstate basis, the W
interaction with quarks can be written as
\begin{eqnarray}
&&\mathcal{L}= -{g\over \sqrt{2}} \bar U \gamma^\mu V_{KM} L D
W^+_\mu + h.c.,\nonumber
\end{eqnarray}
where $V_{KM}$ is an unitary matrix. It is also
called the CKM matrix $V_{CKM}$. For $N$ generations of quark, $V_{CKM}$ in general has $N(N-1)/2$
mixing physical angles and $(N-1)(N-2)/2$ physical phases. A non-zero value for the
sine of the phases lead to CP violation. The minimal number of
generations for CP violation is three.

With three generations of quark, $V_{CKM}$ can be written as
\begin{eqnarray}
V_{CKM} &=& \left ( \begin{array}{ccc} V_{ud}&V_{us}&V_{ub}\\
V_{cd}&V_{cs}&V_{cb}\\
V_{td}&V_{ts}&V_{tb}
\end{array}
\right )\nonumber\\
&=&\left ( \begin{array}{lll}
c_{12}c_{13}&s_{12}c_{13}&s_{13}e^{-i\delta_{13}}\\
-s_{12}c_{23}-c_{12}s_{23}s_{13}e^{i\delta_{13}}&
c_{12}c_{23}-s_{12}s_{23}s_{13}e^{i\delta_{13}}&s_{23}c_{13}\\
s_{12}s_{23}-c_{12}c_{23}s_{13}e^{i\delta_{13}}&
-c_{12}s_{23}-s_{12}c_{23}s_{13}e^{i\delta_{13}}&c_{23}c_{13}
\end{array} \right ),\label{kmmatrix}
\end{eqnarray}
where $s_{ij} = \sin\theta_{ij}$ and $c_{ij} = \cos\theta_{ij}$,
and $\gamma = \delta_{13}$. A convenient parametrization used often is
the Wolfenstein
parametrization\cite{wolfenstein} with
 $V_{us}=s_{12}c_{13}=\lambda$,
$V_{ub}=s_{13}e^{-i\gamma} =A\lambda^3(\rho -i\eta)$, $V_{cb} =
s_{23}c_{13} = A\lambda^2$.

\begin{figure}[!htb]
\begin{center}
%\begin{tabular}{cc}
\includegraphics[width=6cm]{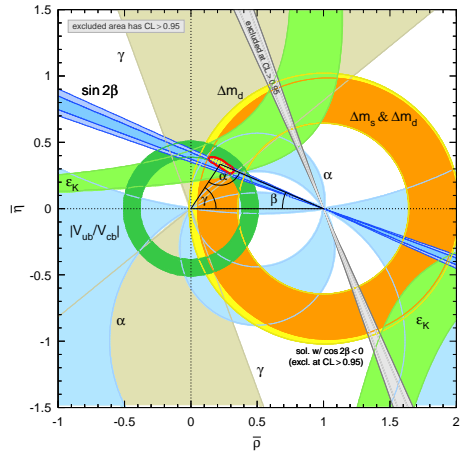}
%\end{tabular}
\end{center}
\label{figure} \caption{Constraints on the CKM parameters.}
\label{diagram}
\end{figure}

The CKM model is very successful in describing all laboratory
experimental results related to CP violation and mixing phenomena.
In Fig. 1 we show the current constraints from various
experiments\cite{pdg}. The best fit values and their error bars for the parameters
are\cite{pdg}
\begin{eqnarray}
\lambda = 0.2272\pm 0.0010\;,\;\;A =
0.818^{+0.007}_{-0.017}\,\;\;\rho =
0.221^{+0.064}_{-0.028}\;,\;\;\eta = 0.340^{+0.017}_{-0.045}\; .
\nonumber
\end{eqnarray}

The CKM model although successful, provides no explanation where
the CP violating phase comes from which calls for more theoretical
studies. There is also hint from matter and anti-matter asymmetry
in our universe that there is the need of CP violation beyond the
CKM model since it gives too small a value for the observed
asymmetry. The origin of CP violation is an outstanding problem of
particle physics.

Also in 1973, T. D. Lee\cite{tdl} proposed that CP violation can
come from symmetry breaking in the vacuum, spontaneous CP
violation. This provides a understanding of the origin of CP
violation. In the minimal SM this is, however, not possible. With
more than one Higgs doublets, it can be realized. For example with
two Higgs doublet $\phi_1$ and $\phi_2$, the most general
potential one can write down is given by\cite{tdl}
\begin{eqnarray}
V(\phi) &=& -\lambda_1\phi^\dagger_1 \phi_1 - \lambda_2
\phi^\dagger_2 \phi_2 - \lambda_{12}(\phi^\dagger_1 \phi_2 +
\phi_2^\dagger \phi_1)\nonumber\\
&& + A(\phi^\dagger_1 \phi_1)^2 + B (\phi^\dagger_2 \phi_2)^2+C
(\phi^\dagger_1\phi_1)(\phi^\dagger_2\phi_2) +\bar
C(\phi^\dagger_1\phi_2)(\phi^\dagger_2\phi_1)  \nonumber \\
&&\, +{\frac{1}{2}}[(\phi^\dagger_1\phi_2)(D\phi^\dagger_1\phi_2 +
E \phi^\dagger_1 \phi_1 + F\phi^\dagger_2\phi_2) + h.c.].
\nonumber
\end{eqnarray}
Writing the VEVs as $\langle \phi^0_1\rangle =
\rho_1e^{i\theta}/\sqrt{2}$, $\langle \phi^0_2\rangle =
\rho_2/\sqrt{2}$, if $\theta \neq 0$ CP is spontaneously violated.
One of the conditions for the Higgs potential to be minimal at the
VEV is
\begin{eqnarray}
\left . {\partial V\over \partial \theta}\right | _{min} = (2
\lambda_{12} - 4 D\cos\theta - E \rho_1^2 - F \rho^2_2)
\rho_1\rho_2 \sin\theta = 0.
\end{eqnarray}
This has a non-trivial solution for spontaneous CP violation given by
\begin{eqnarray} \cos\theta = {1\over 4 D} (2\lambda_{12} -
E\rho_1^2 - F\rho^2_2).
\end{eqnarray}

The above two Higgs doublet model has complicated interactions
with quarks. The Yukawa interactions and mass matrices are given
by
\begin{eqnarray}
&&\mathcal{L}_{\phi-q} = - \bar Q_L (\lambda^u_1 \phi_1 +
\lambda_2^u \phi_2) U_R - \bar Q_L (\lambda^d_1 \tilde \phi_1 +
\lambda_2^d \tilde
\phi_2) D_R,  \nonumber \\
&&M_u = {\frac{1}{\sqrt{2}}}(\lambda^u_1 \rho_1 e^{i\theta} +
\lambda^u_2 \rho_2),\;\; M_d = {\frac{1}{\sqrt{2}}}(\lambda^d_1
\rho_1e^{-i\theta} + \lambda^d_2 \rho_2).  \nonumber
\end{eqnarray}
Even though $\lambda^{u,d}_{1,2}$ are real, because $M_{u,d}$ are
complex one can pbtain a complex $V_{CKM}$. However in such a
model, it is not clear how the spontaneous CP violating phase
$\theta$ is related to the CKM phase $\delta_{13}$. There are also
tree level flavor changing neutral current (FCNC) due to exchange
of neutral Higgs,
\begin{eqnarray}
\mathcal{L}_{neutral} =-{\frac{1}{\sqrt{2}}}[\bar U_L \hat
\lambda^u_2 U_R (h^0_2 - {\frac{\rho_2}{\rho_1}} h^0_1 + i
{\frac{\rho}{\rho_1}} A ) + \bar
D_L \hat \lambda^d_2 D_R (h^0_2 - {\frac{\rho_2}{\rho_1}} h^0_1 - i {\frac{%
\rho}{\rho_1}} A) ],  \nonumber
\end{eqnarray}
$\rho=\rho^2_1+\rho_2^2$, The hat indicates $\lambda^{u,d}_2$ are
in their mass eigenbasis which are in general complex. Higgs
potential will mix the physical Higgs degrees of freedom $A$ and
$h^0_i$.

The model can be made to be consistent with data, but CP violation
are very non-CKM like in general, and Higgs masses are constrained
to be large due to tree level FCNC. To avoid FCNC at tree level,
Weinberg\cite{weinberg1} in 1976 proposed to impose additional
discrete symmetries such that only one Higgs doublet gives masses
to the up and/or down quark sectors. In this case three Higgs
doublets are needed to have spontaneous CP violation. In the
Weiberg model the CKM matrix is real and in conflict with
experimental data\cite{problem}, in particular with data for $\sin
2\beta_{eff} = 0.678\pm0.032$ from $B\to J\psi K$ decay\cite{pdg}.
In the Weinberg model there is no contribution from CKM sector,
and the charged Higgs contribution to $\sin2\beta_{eff}$ is less
than 0.05. The Weinberg model of spontaneous CP violation is
decisively ruled out by CP violation in $B \to J/\psi K$.

In the following we present a model\cite{model} where the CP
violating phase in the CKM matrix has a clear relation with the
spontaneous CP violating phase in the Higgs potential by
identifying these two phases to be the same up to a sign.

We start the discussion by showing that the above idea is indeed
realizable.  Let us consider the following Yukawa couplings with
multi-Higgs doublets,
\begin{eqnarray}
L_Y = \bar Q_L (\Gamma_{u1} \phi_1 + \Gamma_{u2} \phi_2) U_R +
\bar Q_L \Gamma_d \tilde \phi_d D_R + h.c.\;,
\end{eqnarray}
$\tilde \phi_d = -i\sigma_2 \phi_d^*$ and $\phi_d$ may be one of
the $\phi_{1,2}$ or another doublet Higgs field. The Yukawa couplings $%
\Gamma_{u1,u2,d}$ must be real if CP is only violated
spontaneously.

The Higgs doublets when expressed in terms of the component fields
and their VEV $v_i$ are given by
\begin{eqnarray}
\phi_i = e^{i\theta_i}H_i = e^{i\theta_i}\left (
\begin{array}{c}
{\frac{1}{\sqrt{2}}}(v_i +R_i +i A_i) \\
h^-_i
\end{array}
\right ).
\end{eqnarray}
The quark mass terms in the Lagrangian are
\begin{eqnarray}
L_m &= &-\bar U_L \left [M_{u1}e^{i\theta_1} + M_{u2}
e^{i\theta_2}\right ]U_R - \bar D_L M_d e^{-i\theta_d}D_R +
h.c.\;,
\end{eqnarray}
where $M_{ui} =-\Gamma_{ui} v_i / \sqrt{2}$.

The phases $\theta_1$ and $\theta_d$ can be absorbed by redefining
the fields $U_R$ and $D_R$. However, the phase difference $\delta
= \theta_2 - \theta_1$ cannot be removed and it depends on the
Higgs potential. A non-zero $\delta$ indicates spontaneous CP
violation. Without loss of generality, we work in the basis where
$D_L$, $D_R$ are
already in their mass eigenstates. In this basis the down quark mass matrix $%
M_d$ is diagonalized, which will be indicated by $\hat M_d$. In
general the up quark mass matrix $M_u = M_{u1} +
e^{i\delta}M_{u2}$ is not diagonal. Diagonalizing $M_u$ produces
the CKM mixing matrix. One can write $\hat M_u = V_{CKM}
M_uV^\dagger_R$. $V_R$ is an unknown unitary matrix. A direct
identification of the phase $\delta$ with the phase $\delta_{13}$
in the CKM matrix is not possible in general. There are, however,
classes of mass matrices which allow such a connection. A simple
example is provided by setting $V_R$ to be the unit matrix. With
this condition, $M_u = V^\dagger_{CKM} \hat M_u$. One then needs
to show that $V_{CKM}^\dagger$ can be written as
\begin{eqnarray}
V_{CKM}^\dagger = (M_{u1} + e^{i\delta} M_{u2})\hat M^{-1}_u.
\label{mvkm}
\end{eqnarray}
Expressing the CKM matrix in this form is very suggestive. If
$V_{CKM}$  can always be written as a sum of two terms with a
relative phase, then the phase in the CKM matrix can be identified
with the phase $\delta$.

We now demonstrate that it is the case by using the
parametrization in Eq. (\ref{kmmatrix}) as an example.
To get as close as to the form in Eq. (%
\ref{mvkm}), we write the CKM matrix eq. (\ref{kmmatrix})
as\cite{pdg}
\begin{eqnarray}
V_{CKM} = \left(
\begin{array}{lll}
e^{-i\delta_{13}} & 0 & 0 \\
0 & 1 & 0 \\
0 & 0 & 1
\end{array}
\right) \left(
\begin{array}{lll}
c_{12}c_{13}e^{i\delta_{13}} & s_{12}c_{13}e^{i\delta_{13}} & s_{13} \\
-s_{12}c_{23}-c_{12}s_{23}s_{13}e^{i\delta_{13}} &
c_{12}c_{23}-s_{12}s_{23}s_{13} e^{i\delta_{13}} & s_{23}c_{13} \\
s_{12}s_{23}-c_{12}c_{23}s_{13}e^{i\delta_{13}} &
-c_{12}s_{23}-s_{12}c_{23}s_{13} e^{i\delta_{13}} & c_{23}c_{13}
\end{array}
\right ).\nonumber
\end{eqnarray}

Absorbing the left matrix into the definition of $U_L$ field, we
have
\begin{eqnarray}
&&M_{u1} = \left (
\begin{array}{ccc}
0 & - s_{12}c_{23} & s_{12}s_{23} \\
0 & c_{12}c_{23} & -c_{12}s_{23} \\
s_{13} & s_{23}c_{13} & c_{23}c_{13}
\end{array}
\right ) \hat M_u\;,\nonumber\\
&& M_{u2} = \left (
\begin{array}{ccc}
c_{12}c_{13} & -c_{12}s_{23}s_{13} & -c_{12}c_{23}s_{13} \\
s_{12}c_{13} & -s_{12}s_{23}s_{13} & -s_{12}c_{23}s_{13} \\
0 & 0 & 0
\end{array}
\right ) \hat M_u\;,
\end{eqnarray}
and $\delta = -\delta_{13}$. We therefore find that it is possible
to identify the CKM phase with that resulting from spontaneous CP
violation. Note that as long as the phase $\delta$ is not zero, CP
violation will show up in the charged currents mediated by W
exchange. The effects do not disappear even when Higgs boson
masses are all set to be much higher than the $W$ scale.
Furthermore, $M_{1,2}$ are fixed in terms of the CKM matrix
elements and the quark masses, as opposed to being arbitrary in
general multi-Higgs models.

We comment that the solution is not unique even when $V_R$ is set
to be the unit matrix\cite{model}. One can easily verify this by
taking another parametrization for the CKM matrix, such as the
original Kobayashi-Maskawa (KM) matrix\cite{km}. More physical
requirements are needed to uniquely determine the connection.  The
key point we would like to emphasis is that there are solutions
where the phase in the CKM matrix can be identified with the phase
causing spontaneous CP violation in the Higgs potential.

The mass matrices $M_{u1}$ and $M_{u2}$ can be written in a more
elegant way with
\begin{eqnarray}
M_{u1}&=& V^\dagger_{CKM}\hat M_u -
{\frac{e^{i\delta}}{\sin\delta}}
Im(V^\dagger_{CKM})\hat M_u\;,  \nonumber \\
M_{u2}&=& {\frac{1}{\sin\delta}} Im(V^\dagger_{CKM}) \hat M_u\;.
\end{eqnarray}

Alternatively, a model can be constructed with two Higgs doublets
couple to the down sector and one Higgs doublet couples to the up
sector\cite{model}. We will concentrate on the above scenario for
detailed discussions. In the following we go further to construct a realistic model.

A common problem
for models with spontaneous CP violation is that a strong QCD
$\theta$ term will be generated\cite{bigi}. Constraint from
neutron dipole moment measurement will rule out spontaneous CP
violation as the sole source if there is no mechanism to make sure
that the $\theta$ term is small enough. The model mentioned above
faces the same problem. We therefore supplement the model with a
Peccei-Quinn (PQ) symmetry\cite{pq} to ensure a small $\theta$.

To have spontaneous CP violation and also PQ symmetry
simultaneously, more than two Higgs doublets are
needed\cite{hvgn}. For our purpose we find that in order to have
spontaneous CP violation with PQ symmetry at least three
Higgs doublets $\phi_i = e^{i\theta_i}H_i$ and one complex Higgs singlet $%
\tilde S = e^{i\theta_s}S = e^{i\theta_s}(v_s + R_s + i
A_s)/\sqrt{2}$ are required. The Higgs singlet with a large VEV
renders the axion from PQ symmetry breaking to be invisible\cite
{invisible,kk}, thus satisfying experimental constraints on axion
couplings to fermions. We will henceforth work with models with an
invisible axion\cite{invisible} with PQ charges for various fields
given by
\begin{eqnarray}
&&Q_L : 0\;,\;\;U_R: -1\;,\;\;D_R: -1\;,\;\;\phi_{1,2}:
+1\;,\;\;\phi_d=\phi_3: -1\;,\;\;\tilde S:+2\;. \nonumber
\end{eqnarray}

The Higgs potential is
\begin{eqnarray}
V&=& -m^2_1 H_1^\dagger H_1-m^2_2 H_2^\dagger H_2-m^2_3
H_3^\dagger H_3-m^2_{12} (H_1^\dagger H_2e^{i(\theta_2 -
\theta_1)}+ h.c.)-m^2_s
S^\dagger S  \nonumber \\
&+&\lambda_1 (H_1^\dagger H_1)^2+\lambda_2 (H_2^\dagger H_2)^2
+\lambda_t
(H_3^\dagger H_3)^2 + \lambda_s (S^\dagger S)^2  \nonumber \\
&+&\lambda_3 (H_1^\dagger H_1)(H^\dagger_2H_2)+ \lambda^{\prime}_3
(H_1^\dagger H_1)(H^\dagger_3H_3)+\lambda^{\prime\prime}_3
(H_2^\dagger
H_2)(H^\dagger_3H_3)  \nonumber \\
&+&\lambda_4 (H_1^\dagger H_2)(H^\dagger_2H_1)+ \lambda^{\prime}_4
(H_1^\dagger H_3)(H^\dagger_3H_1)+\lambda^{\prime\prime}_4
(H_2^\dagger
H_3)(H^\dagger_3H_2)  \nonumber \\
&+&{\frac{1}{2}}\lambda_5
((H^\dagger_1H_2)^2e^{i2(\theta_2-\theta_1)}+h.c.) +\lambda_6
(H^\dagger_1 H_1)(H^\dagger_1 H_2 e^{i(\theta_2-\theta_1)}+h.c.)
\nonumber \\
&+&\lambda_7 (H^\dagger_2 H_2)(H^\dagger_1 H_2
e^{i(\theta_2-\theta_1)}+h.c.)+\lambda_8 (H^\dagger_3
H_3)(H^\dagger_1 H_2
e^{i(\theta_2-\theta_1)}+h.c.)  \nonumber \\
&+& f_1 H^\dagger_1 H_1 S^\dagger S + f_2 H^\dagger_2 H_2
S^\dagger S + f_3 H^\dagger_3 H_3 S^\dagger S\nonumber\\
& +& d_{12} (H^\dagger_1 H_2e^{i(\theta_2-\theta_1)} + H^\dagger_2
H_1e^{-i(\theta_2 -\theta_1)})
S^\dagger S  \nonumber \\
&+& f_{13} (H^\dagger_1 H_3 S e^{i(\theta_3 + \theta_s -
\theta_1)} + h.c.)+f_{23} (H^\dagger_2 H_3 S e^{i(\theta_3 +
\theta_s - \theta_2)} + h.c.)\;.
\end{eqnarray}
Note that only two independent phases occur in the above
expression, which we choose to be $\delta = \theta_2-\theta_1$ and
$\delta_s = \theta_3 +\theta_s - \theta_2$. The phase
$\theta_3+\theta_s - \theta_1$ can be written as $\delta +
\delta_s$. Differentiating with respect to $\delta_s$ to get one
of the conditions for minimization of the potential, we get
\begin{eqnarray}
&&f_{13}v_1v_3 v_s \sin(\delta_s + \delta) + f_{23} v_2v_3 v_s
\sin\delta_s = 0\;.
\end{eqnarray}
It is clear that $\delta$ and $\delta_s$ are related with
\begin{eqnarray}
\tan\delta_s = -{\frac{f_{13} v_1 \sin\delta}{f_{23}v_2 + f_{13}
v_1 \cos\delta}}\;.
\end{eqnarray}

The phase $\delta$ is the only independent phase in the Higgs
potential. A non-zero $\sin\delta$ is the source of spontaneous CP
violation and also the only source of CP violation in the model.

Removing the would-be Goldstaone bosons, one can write the Yukawa
interactions for physical Higgs degrees of freedom as the
following\cite{model}
\begin{eqnarray}
L_Y &=& \bar U_L [{\hat M_{u} }\frac{v_1}{v_{12}v_2} - (\hat
M_{u}-V_{CKM}Im(V^\dagger_{CKM})\hat M_u
{\frac{e^{i\delta}}{\sin\delta}} )
\frac{v_{12}}{v_1v_2} ]U_R(H_1^0+i a_1^0)  \nonumber \\
&+& \bar U_L {\hat M_{u}} U_R [\frac{v_3}{v_{12}v}(H_2^0+i a_2) -\frac{1}{v}%
H_3^0+\frac{v_3^2}{v^2 v_s}(H_4^0 + i a )]  \nonumber \\
& -& \bar D_L \hat M_d D_R
[\frac{v_{12}}{v_3v}(H_2^0-ia_2)+\frac{1}{v}H_3^0
+\frac{v_{12}^2}{v^2 v_s}(H_4^0 -i a) ]  \nonumber \\
&+& \sqrt{2} \bar D_L [V_{CKM}^\dagger \hat
M_{u}\frac{v_1}{v_2v_{12}}
-(V_{CKM}^\dagger \hat M_{u}- Im(V_{CKM}^\dagger)\hat M_u {\frac{e^{i\delta}%
}{\sin\delta}})\ \frac{v_{12}}{v_1v_2}]U_R {H^-_1}  \nonumber \\
&-&\sqrt{2}\frac{v_3}{v_{12}v} \bar D_L V_{CKM}^\dagger {\hat M_u}
U_R H^-_2 - \sqrt{2}\frac{v_{12}}{vv_3} \bar U_L V_{CKM} \hat M_d
D_R H_2^+ + h.c.\;.
\end{eqnarray}
Here $a$ is the axion. The fields, $H^0_i$ and $a^0_i$ are not the
mass eigenstates yet and will be mixied in the potential.

Note that the couplings of $a$ and $H^0_4$ to quarks are
suppressed by $1/v_s $, and that only the exchange of $H_1^0$ and
$a_1^0$ can induce tree level
FCNC interactions. The FCNC coupling is proportional to $%
V_{CKM}Im(V_{CKM}^\dagger)\hat M_u$.

For the model presented here, FCNC only involves the up quark
sector. The most stringent constraint on the Higgs mass comes from
$D^0-\bar D^0$. We find that the Higgs mass can be as low as a
hundred GeV from this constraint. Such low Higgs mass can be
probed at LHC and ILC.

The neutron EDM $d_n$ provides much information on the model
parameters. The standard model predicts a very small\cite{hmp}
$d_n$ ($<10^{-31}e$ cm). The present experimental upper bound on
neutron EDM $d_n$ is very tight\cite{pdg}: $|d_n|<0.63\times
10^{-25} e$ cm. In the model considered above, the quark EDMs will
be generated at loop levels due to mixing between $a_i$ and $H_i$.

The one loop contributions to the neutron EDM are suppressed for
the usual reason of being proportional to light quarks masses to
the third power for diagram in which the internal quark is the
same as the external quark. In our model, there is a potentially
large contribution when there is a top quark in the loop. However,
the couplings to top are proportional to $s_{13}$, therefore the
contribution to neutron EDM is much smaller than the present upper
bound. It is well known that exchange of Higgs at the two loop
level may be more important than the one loop contribution,
through the quark EDM, quark color EDM, and the gluon color EDM.
We find that in the model discussed above, the neutron EDM can
reach the present experimental bound. Improved measurement on
neutron EDM can provide us with more information.

To summarize, I have described a model in which the CP violating
phase in the CKM mixing matrix to be the same as that causing
spontaneous CP violation in the Higgs potential. When the Higgs
boson masses are set to be very large, the phase in the CKM matrix
can be made finite and CP violating effects will not disappear. An
interesting feature of this model is that the FCNC Yukawa
couplings are fixed in terms of the quark masses and CKM mixing
angles, making phenomenological analysis much easier.

\noindent{\bf Acknowledgments} This work was supported in part by
NSC and NCTS. I thank S.-L. Chen, N.G. Deshpande, J. Jiang and
L.-H. Tsai for collaboration on the work reported here.

%\tighten

\end{document}